\documentclass[journal]{IEEEtran}
\usepackage{mwe} 
\usepackage[nopar]{kantlipsum} 
\usepackage{amsmath}
\usepackage{amssymb}
\usepackage{color,soul}
\usepackage{graphicx}
\usepackage{epstopdf}
\usepackage[noend]{algpseudocode}
\usepackage{algorithm2e}
\usepackage{multirow}
\usepackage{tabularx}
\usepackage{cite}
\usepackage{hyperref}
\graphicspath{ {figures/} }

\begin{document}
\title{Cascading Failure Mitigation via Transmission Switching}

\author{Sayed Abdullah Sadat,~\IEEEmembership{Graduate Student Member,~IEEE,}
        and~Mostafa~Sahraei-Ardakani,~\IEEEmembership{Member,~IEEE,}
\thanks{Authors are with the Department
of Electrical and Computer Engineering, University of Utah, Salt Lake City,
UT, 84112, USA. (e-mail: sayed\_abdullah@ieee.org; mostafa.ardakani@utah.edu).}}

\markboth{Submitted for Review}%
{Shell \MakeLowercase{\textit{et al.}}: Bare Demo of IEEEtran.cls for IEEE Journals}

\maketitle

\begin{abstract}
After decades of research, cascading blackouts remain one of the unresolved challenges in the bulk power system operations. A new perspective for measuring the susceptibility of the system to cascading failures is clearly needed. The newly developed concept of system stress metrics may be able to provide new insights into this problem. The method measures stress as susceptibility to cascading failures by analyzing the network structure and electrical properties. This paper investigates the effectiveness of transmission switching in reducing the risk of cascading failures, measured in system stress metrics. Based on line-outage distribution factors, an algorithm is developed to identify and test the switching candidates quickly.
A case study analyzing different stress metrics on the IEEE 118-bus test system is presented. The results show that transmission switching identified by our proposed algorithm could be used in preventive as well as corrective mechanisms to reduce the system's susceptibility to cascading failures. Contrary to the conventional operation wisdom that switching lines out of service jeopardizes reliability, our results suggest the opposite; system operators can often use transmission switching, when the system is under stress, as a tool to reduce the risk of cascading failures. 
\end{abstract}


%
\IEEEpeerreviewmaketitle

\section{Introduction}
%
%
%
%
\IEEEPARstart{T}{here} is an extensive body of academic and industry literature on analyzing cascading blackouts, seeking ways to eliminate them or at least reduce their frequency and size, and improve the speed of recovery~\cite{merrill}. 
Most of the blackouts have been subject to investigations and postmortem analyses~\cite{merrill}.
The largest blackout in the North American grid, the Northeast blackout of 2003, was studied for over a year. The results of this extensive analysis were published in an illuminating three-volume report~\cite{2003blackout}. This report also provides useful insight into several earlier cascading failure events~\cite{merrill}. The North American Electric Reliability Corporation (NERC) was established by the US-Canadian power industry after the 1965 blackout to improve reliability, notably by producing criteria and collecting data. Preventing cascading blackouts has always been central to the objectives of these criteria~\cite{TPL4}. In 1974, state estimation was introduced in power systems to have more accurate inputs to real-time procedures for increasing reliability. In the context of cascading failures, the purpose of state estimation was more to make data availability reliable rather than improving the data accuracy~\cite{merrill5}. Before that, a state model was developed in which necessary considerations for designing a total control system for reliability improvement of the generation and transmission systems were incorporated. In this model, the control system was made of automatic functions, human participation, and an information system~\cite{merrill4}. Much labor has been invested in a host of efforts to solve the blackout problem. Recently, network theory has been applied to blackouts and other power systems issues, but blackouts continue; the problem has not been solved~\cite{merrill}.

\subsection{Cascading Failures}

Cascading failures in large systems can be due to at least one of the following reasons:~\cite{merrill}
\begin{enumerate}
\item Failure of protection system and control devices;
\item Failure of processes and procedures;
\item Overly stressed loading scenarios;
\end{enumerate}

\begin{figure*}[h!]
	\centering
	\includegraphics[width=5.1in]{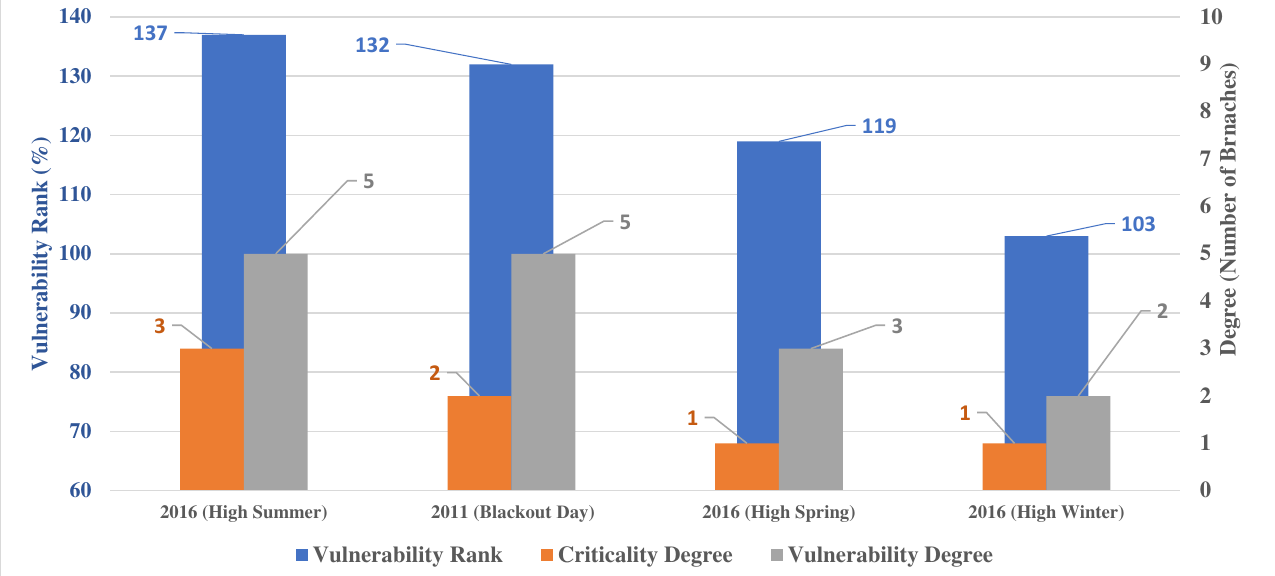}
	\caption{Comparison of the stress on Western interconnection for the peak load during different seasons in 2016 and the blackout day in September 2011 \cite{abid}.} \label{fig:abids}
\end{figure*}

Among these failure causes, the possibilities of forestalling the first two (i.e., ``control and protective devices" and ``policies and procedures") or even testing for these failures are astronomical and individual events are improbable. We simply lack models that would reflect these two's effects on the system because we do not know if or how a failure will lead to a cascading failure. In other words, we cannot model protection and control system failures and failures in processes and procedures into our bulk electric system model. However, history shows that cascading also depends critically on how the system is loaded, which can be described by the system stress metrics~\cite{merrill, sayed3}. This trend can be observed in the 2011 Western Interconnection post blackout study, shown in Fig.~\ref{fig:abids}. The figure shows the system stress for four different loading scenarios, with three stress metrics. Arizona and Southern California are most stressed during the peak in the summer, as the system is heavily loaded. The system is usually not nearly as stressed during spring and winter peaks. The Southwest blackout of 2011 occurred on September 8 at around 3:38 PM PDT. This time is not usually regarded as a peak hour; however, as shown in Fig.~\ref{fig:abids}, stress metrics reveal that the system was indeed atypically stressed before the blackout. The blackout was initiated by a technician mistake, who switched a 500 kV line between APS's Hassayampa and North Gila substations in Arizona~\cite{sandiego}. This blackout could have been avoided if the stress was identified and reduced~\cite{abid}.

Several metrics have been developed to capture grid operation risks \cite{mccalley,overbye,merrill}. These metrics have their pros and cons. Multiple element contingency consideration provides more levels of visibility of stress on the stress as in \cite{overbye}. Other metrics take into consideration the voltage profile of the system and the probability of the failures of transmission elements \cite{mccalley}. We believe that it has to be computationally efficient and inexpensive for metrics to offer real-time stress visibility, preferably employing the existing Energy Management System (EMS) tools. Considering multiple-element contingencies are computationally very expensive. The majority of the computationally less expensive algorithms for metrics often capture less than one-third of the sever contingencies, which are more likely to be followed by a cascading failure, thereby making these metrics less effective \cite{overbye}.
Besides, cascading failures that began with N-2 (i.e., simultaneous independent contingencies instead of causal N-1-1 contingencies, where a first contingency caused a second to occur) are rarely heard of. We believe that considering the voltage factor in the metrics, which brings in significant additional computational expense are less helpful. Ignoring the voltage impact does not impact the viability of our method. It can be verified by another work \cite{xli}, which analyzed transmission switching in the PJM network, one of the largest transmission network operators globally, has shown that more than 99\% of the switching transmission elements provide a stable solution as expected by the NERC standards. Additionally, the response to voltage disturbances by the protection system is usually slow compared to its response to overload, which is instantaneous. The slow response to voltage change provides enough time for the operator to mitigate the impact and take corrective actions. We also prefer to avoid probabilistic methods of approach to produce metrics simply because of the varieties of factors that can contribute to the probability of a contingency and, consequently, the computational burden of it versus the benefit it can offer to an operator in real-time. A set of new tools, including metrics of stress or susceptibility to cascading failures, which were introduced and discussed in \cite{merrill, sayed3}, have been adopted and further extended. We found these tools more appropriate for determining the potential elements for transmission switching in real-time operations. They have been built on two very different theoretic bases to develop methods that planners and operators can use to spot stressed areas in different operating states and subsequently plan and operate the system securely. The tool has been successfully implemented on Peru System, Eastern Interconnection, and Western Interconnection to explore the system's susceptibility to cascading failures~\cite{merrill_conv}. In every instance of cascading blackouts, the stress metrics were able to show unusual system stress before the event~\cite{merrill_conv}.

\subsection{Transmission Switching}
Transmission switching (TS) refers to changing the transmission network's topology by opening or closing transmission lines. The concept, which was first introduced in 1968 by the German mathematician Dietrich Braess, is counter-intuitive but a well-known fact that removing edges from a network with “selfish routing” can decrease the latency incurred by traffic in an equilibrium flow. Since then, a large body of academic literature has been dedicated to studying this paradox in infrastructure networks~\cite{braessparadox}. 
This concept was first proposed in the power systems in the 1980s; in the following years, several studies adopted transmission switching as a corrective mechanism~\cite{rolim}. Later, the concept of optimal transmission switching was proposed to minimize the operation cost~\cite{sang}. This technique has recently been integrated within different power systems operation models, such as security-constrained economic dispatch, security-constrained unit commitment, and real-time contingency analysis. TS has been proven to significantly reduce the operational costs and improve system reliability~\cite{sang, goldis, hedman, mojdeh}. Due to computational complexity and other concerns such as dynamic stability, industry adoption of TS has been very limited. Some system operators use TS as a corrective mechanism for improving voltage profiles and mitigating line overloading \cite{californiaiso, neiso}. TS is also being employed during planned outages to make the transition smooth, and also as a post-contingency corrective action \cite{xli}. California ISO (CAISO) is reported to perform TS on a seasonal basis and relieve congestion in the system~\cite{hedman, californiaiso2}. PJM has posted a list of potential switching solutions that may reduce or eliminate violations for normal and post-contingency situations~\cite{PJM, xli2}. However, these switching actions are not always guaranteed to provide benefits because they are identified offline.

The use of transmission switching has been extensively studied for different purposes in power systems. Still, no study yet explicitly looked into the impact of transmission switching on reducing the system's susceptibility to cascading failures.

First, this paper aims to quantify the system's susceptibility to cascading failures in terms of the system's stress. The paper then studies the impacts of transmission switching on reducing the system stress, and as a result, lowering the system's susceptibility to cascading failures. The contributions of this paper can be summarized as follows: 

\begin{enumerate}
\item Further development of statistical metrics to measure system stress by introducing four new metrics, including indices for potential candidates for transmission switching;
\item Examination of the impacts of preventive transmission switching on reducing stress during unusually stressed or poorly forecasted loading scenarios;
\item Investigation of the benefits of corrective transmission switching in lowering the system stress, after N-1 contingencies.

\end{enumerate}

The rest of the paper is organized as follows. In Section II, we introduce stress metrics to measure the susceptibility of the system to cascading failures. Section III presents preventive and corrective transmission switching. Section IV demonstrates the effectiveness of the method via simulation studies on the IEEE 118-bus test system. Finally, Section V concludes this paper.

\section{Measure of Stress} \label{sec:StressMetrics}
Elements from network theory and traditional power system analysis have been combined to measure the system's stress. Consequently, indices are developed, describing how a failure would propagate through a system ~\cite{merrill, sayed3}.

Flow violations on all the lines after every potential contingency are needed to calculate the stress metrics. Such information is readily available from the contingency analysis tool, which is a part of the EMS \cite{xli,sahraei2016real}. Even without access to such information, post-contingency flows can be approximately calculated via line outage distribution factor (LODF), which colloquially is also referred to as DFAX~\cite{merrill, sayed3}. These sensitivities can be computed using conventional power flow software with DC approximation or using the current-based generalized injection shift factors~\cite{ACDFAX}. 

An $LODF_{ij}$ of 0.5 indicates that 50\% of the pre-outage flow on line $j$ would be added to the flow on line $i$, should line $j$ go out of service. Post-outage flow on line $i$ after the outage of line $j$ is calculated in (\ref{eq:LODF_Flow}), where $f^0$ indicates the pre-outage flow.
\begin{align}
    f_i \cong f_{i}^0 + LODF_{ij} \times f_{j}^0\label{eq:LODF_Flow}
\end{align}

This relationship is approximate because the power system is only approximately linear. However, the accuracy of (\ref{eq:LODF_Flow}) is acceptable, and power system planners and operators extensively use  LODFs in contingency analysis~\cite{merrill}. The experience shows that for real-time power analysis, which is believed to be the key issue, the nonlinearity is rarely troublesome~\cite{merrill,sayed2}. It can be argued that nonlinear and dynamic issues, as well as voltage problems, which are not reflected in the linearized LODF, are part of cascading failures. Of course, we agree that such effects occur, for example, in the two famous blackouts we described above. However, cascading failures always begin with the linearizable real-power stresses that our model captures.

The LODF matrix is not symmetrical. However, for a large network, most of its values are rather small. Small values beyond a threshold can be ignored to generate a sparse matrix. The sparsity can, then, be exploited for enhancement of the computation. In a passive linear network, the value of each LODF is between -1.0 and + 1.0. Large positive or negative values of LODF make cascading failure more likely. Tighter coupling is more likely to overload line $i$ and force it out of the service if line $j$ experiences an outage, all else being equal. With an LODF of zero, the outage of the line $j$ by itself will not cause an overload or outage of the line $i$~\cite{merrill,sayed3}.

Network theory suggests analyzing a network with metrics. The metrics we describe below are variants of metrics used commonly in network analysis. The stress metrics proposed in~\cite{merrill,sayed3} reflect the pre-contingency and post-contingency flows. The demand and the generation dispatch determine this pre-contingency loading. These values are hypothetical in planning models; however, in operation, the pre-contingency loading is obtained from real-time metering, which is processed by the state estimator~\cite{sayed3,merrill}. The definition of some of the metrics developed previously and those proposed in this paper are given below.

\subsection{Vulnerability}
Vulnerability deals with the post-outage flow on a monitored line or transformer after the outage of another line or transformer in the system. This is a reasonable measure of stress because cascading failures always begin with an outage, overloading one or more other line or transformer. Consequently, the protection relays will isolate the newly overloaded lines, which will further weaken the system. Two metrics were proposed in \cite{merrill,sayed3} to quantify the vulnerability: the rank and the degree of vulnerability. In addition, we also introduce a new metric for indexing the entire system's vulnerability.

\subsubsection{Rank of Vulnerability ($V_i^{rank}$)}
The rank of vulnerability is the maximum absolute value of flow on a line or transformer per unit of its rating after the outage of another line or transformer. The vulnerability matrix's rank is a $1\times m_1$ matrix, where $m_1$ is the number of monitored lines and transformers. The $i^{th}$ rank of vulnerability is the maximum post-outage flow in the line or transformer $i$ after the outage of all $m_2$ lines and transformers, taken out one at a time, where $m_2$ is the number of lines and transformers, whose outage is monitored. Note that the $i^{th}$ rank of vulnerability may be greater than, less than, or equal to the pre-contingency flow on the line or transformer~\cite{sayed3,merrill}. This metric is expressed as a percentage of the post-contingency flow compared to the line/transformer rating.
 \begin{align}
    V_i^{rank} = \mathrm{max}(\frac{|f_i|}{f_i^{rated}})
\end{align}
\subsubsection{Degree of Vulnerability ($V_i^{degree}$)}
The degree of vulnerability is the number of single outages for which a monitored line or transformer will be loaded over some threshold value. The line's rating is used to compute the degree of vulnerability in this paper. The degree of vulnerability matrix is a $1\times m_1$ matrix, where $m_1$ is the number of monitored lines and transformers. The $i^{th}$ element of this matrix is the number of lines and transformers, among all the $m_2$ lines and transformers, whose outage leads to a power flow beyond the specified threshold for the $i^{th}$ line or transformer.
This metric is calculated and shown in (\ref{eq:VDegree}).
 \begin{align}
    V_i^{degree} = count\_if(\frac{|f_i|}{f_i^{rated}}> Threshold_i)\label{eq:VDegree}
\end{align}

\subsubsection{System Vulnerability Degree ($V_{System}$)}
The system vulnerability rank and degree, proposed in this paper, are respectively the maximum rank of vulnerability and maximum degree of vulnerability of non-radial monitored vulnerable elements. Here, the lines' normal ratings were used to compute the vulnerability number in this study. The system vulnerability metrics are scalar, measured as indices for the entire system, rather than a specific line or transformer.

\subsection{Criticality}
Criticality measures how the outage of a line or transformer affects other lines and transformers in the system. Rank and degree of criticality are used to define criticality~\cite{sayed3,abid,merrill}. In addition, this paper introduces a new metric for measuring the entire system's criticality level.
\subsubsection{Rank of Criticality ($C_i^{rank}$)}
The rank of the criticality of a line or transformer $i$ is the maximum absolute value of flow through all other lines and transformers, per unit of their capacity, after the outage of line or transformer $i$. The rank of criticality matrix is a $1\times m_2$ matrix, where $m_2$ is the number of lines and transformers whose outage is monitored. The $i^{th}$ rank of criticality is the maximum absolute value of all the post-outage flows divided by the ratings of the $m_1$ monitored lines and transformers after the outage of line or transformer $i$~\cite{abid}. This metric is expressed as a percentage of the monitored lines or transformers' rating, as shown in (\ref{eq:CRank}).
 \begin{align}
    C_i^{rank} = \underset{k} {\mathrm{max}} (\frac{|f_k|}{f_k^{rated}})\label{eq:CRank}
\end{align}
\subsubsection{Degree of Criticality ($C_i^{degree}$)}
The degree of criticality of a line or transformer $i$ is the number of monitored lines and transformers that will be loaded above some threshold after the outage of line or transformer $i$. The lines' nominal rating was used here for calculating the degree of criticality, similar to the degree of vulnerability. However, the operator can pick any desirable threshold, and the method does not limit this choice. The degree of criticality matrix is also a $1\times m_2$ matrix, where $m_2$ is the number of lines and transformers whose outage is monitored. The $i^{th}$ degree of criticality is the number of lines and transformers among all the $m_1$ monitored lines and transformers whose flows will exceed the threshold after the outage of the $i_{th}$ line or transformer. This metric can be calculated, as shown in (\ref{eq:CDegree}).
 \begin{align}
    C_i^{degree} = count\_if(\frac{|f_k|}{f_k^{rated}}> Threshold_k)\label{eq:CDegree}
\end{align}

\subsubsection{System Criticality ($C_{System}$)}
The system criticality rank and degree proposed in this paper are the maximum rank of criticality and the maximum degree of criticality of non-radial critical elements, respectively. Nominal line ratings were used to compute the number of criticality in this study. Like the system vulnerability metrics, system criticality metrics are also a scalar measured for the entire system.
\subsection{Switching Stress Relief Index ($SSR$)}
The $SSR^i$ is a metric intended for non-critical lines $i$ to represent their impact on the system if switched. Like other metrics, these metrics are of two types: Switching Stress Relief Rank $SSR^i_r$ and Switching Stress Relief Degree $SSR^i_d$, and are calculated as shown in the algorithm \ref{alg:SSR}.
\begin{algorithm}
\caption{Calculating $SSR^{nc}_d$ and $SSR^{nc}_r$ indices for noncritical lines.}
   \For{$c$ in $Critical\_List$}{
     \For{$nc$ in $Noncritical\_List$}{
        \If {$(F_c^0 < 0  \And LODF_i^{nc} \times F_{nc}^o >0)  \parallel$
        $(F_c^0 > 0 \And LODF_i^{nc} \times F_{nc}^o<0)$}{
        $SSR^{nc}_d \gets SSR^{nc}_d + 1$ \;
        $SSR^{nc}_r \gets SSR^{nc}_r + (|F_c^0|-|F_c|)\times \frac{100}{F_c^{max}}$ \;
        }
     }
   }
   \label{alg:SSR}
\end{algorithm}

The higher the value of $SSR$ of any element, the more preferred candidate is the element for transmission switching to reduce the system's stress.

\section{Transmission Switching}

The electric transmission network is built redundantly to ensure mandatory reliability standards, requiring protection against worst-case scenarios. Due to loop flows in this redundant meshed network, transmission switching may lead to improved economic efficiency and reliability ~\cite{hedman}. This phenomenon is widely acknowledged; however, finding appropriate switching candidates within the available computational time for power system operation remains challenging.

Although transmission switching has many applications, it can be solely performed to enhance the system reliability \cite{sahraei2016real,xli,xli2,minstress}. Reliability-motivated switching is perhaps the first application of transmission switching that is used by the industry \cite{PJM}. References ~\cite{sahraei2016real,xli,xli2} employ transmission switching to reduce the post-contingency network violations. The method proposed in this paper also aims to enhance reliability. We focus on reducing the system stress measured via the metrics Section \ref{sec:StressMetrics}, rather than the post-contingency violation reduction. As mentioned before, transmission switching is considered to be a computationally challenging problem. A recent method, which achieved tractability for reliability-motivated switching, handled this challenge by only allowing a minimal set of switching candidates \cite{sahraei2016real,xli,xli2}. The switchable elements were picked either from a small vicinity of the contingency or the violation. The extensive analysis showed that this small subset includes almost all quality solutions \cite{sahraei2016real,xli,xli2}. In this paper, we use a similar approach by only relying on a small subset of switchable lines; however, rather than searching within the vicinity of contingency or violation, we employ the LODF matrix to choose the most effective switching candidates. The potential candidates can be selected by looking at the overloaded line column corresponding to a contingency. A high negative LODF value is one of the indications of the likely line for switching. Thus, the LODF matrix will provide us with a smart and fast method to select the switching candidates.

Switching is generally classified into two categories, depending on its timeline: preventive and corrective transmission switching. We consider both of these categories in the next two subsections, in the context of system stress reduction.

\subsection{Preventive Transmission Switching}
Preventive action in power system operation is taken to avoid the adverse consequences of a potential disturbance. The disturbance may never happen, but the preventative measure will protect the system against it, should it actually occur. In this paper, we use preventive transmission switching to reduce the system's susceptibility to cascading failures. History shows that cascading failures depend critically on how the system is loaded, which can be described by the system stress metrics. Post-blackout investigations show that, in most cases, the system was atypically stressed before the blackout~\cite{merrill}. Had the stress of the system been taken care of, the blackout could have been prevented. This can be seen in~\cite{abid}, where the stress in the San Diego area was analyzed over different seasons of the year and found that on September 8, 2011, and before the blackout that happened later on the same day, the system was atypically stressed. The blackout could have been avoided through appropriate preventive actions that reduce the system stress.

\begin{figure}[h!]
	\centering
	\includegraphics[width=3.2in]{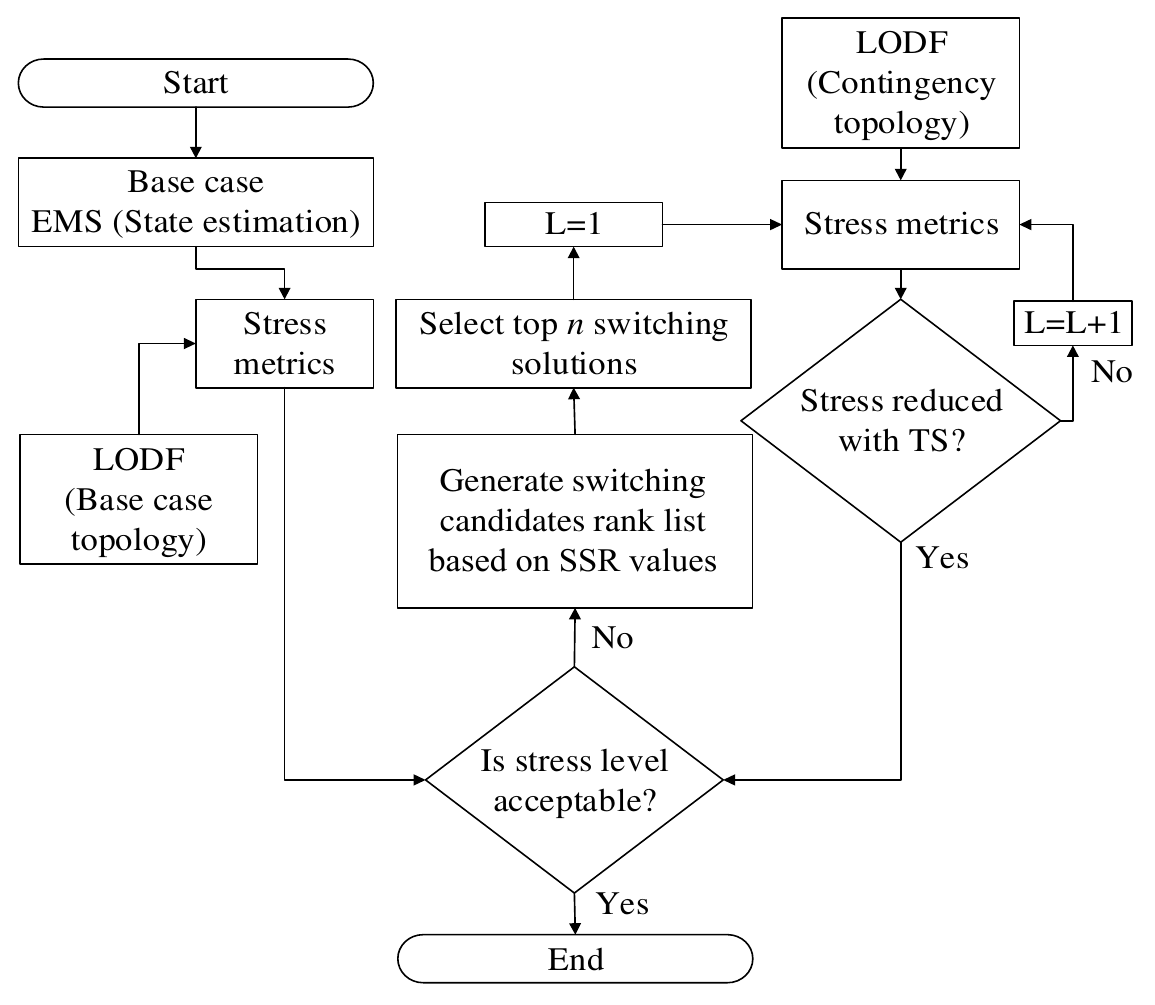}
	\caption{The algorithm, proposed in this paper, to reduce the system stress using transmission switching.} \label{fig:FlowChart}
\end{figure}

This paper proposes that preventive transmission switching should be looked at whenever the system is under atypical stress, beyond a predefined level, to reduce the stress on the system. We hypothesize that transmission switching may offer a cheap and fast solution, relieve the system stress, and avoid potential cascading failures. Once the system stress has been reduced due to a change in the loading, the line can be switched back if it provides economic benefits. Other alternatives can be implemented as preventive actions, such as generation redispatch. However, transmission switching can be implemented much faster and is often the cheapest option, as it only involves the operation of a circuit breaker.
Moreover, generation redispatch is often depleted during the stressed operating scenarios and not available anymore to the operator. Fig.~\ref{fig:FlowChart} shows the proposed algorithm for transmission switching in response to atypical stress on the system. An example of an atypically stressed system was shown in Fig.~\ref{fig:abids}, which led to the Southwest blackout of 2011.

\subsection{Corrective Transmission Switching (CTS)}

 As transmission switching can be implemented instantaneously, unlike generation redispatch, which is relatively slow, more studies have focused on corrective transmission switching than preventive transmission switching. Corrective transmission switching solutions are identified beforehand, within the contingency analysis tool, and are ready for implementation \cite{xli, sahraei2016real}. Only after the contingency occurs does the operator need to implement the solution.

This paper aims to study how corrective transmission switching can reduce post-contingency stress rather than the post-contingency violations. The two are related but are not the same. This paper examines two hypotheses regarding corrective transmission switching. First, we analyze the post contingency system stress imposed on the system by the possibility of an N-1-1 event for the system that is already in the N-1  state. The second point of interest for us is to monitor the ongoing stress level in the system, which is measured in terms of the lines that have already exceeded their contingency limits. These overflows should be addressed within a short period, defined by the emergency limit's maximum duration; otherwise, the overloaded lines may trip and initiate a cascading failure. Thus, transmission switching can either be considered corrective concerning the current post-N-1 state or preventive concerning the possibility of an N-1-1 event.

\begin{table*}
	\centering
	\caption{The criticality and vulnerability stress metrics for the IEEE 118 bus test case at 97\%, 105\%, 110\% of peak loading.} \label{fig:table_97}
		\includegraphics[width=5.0in]{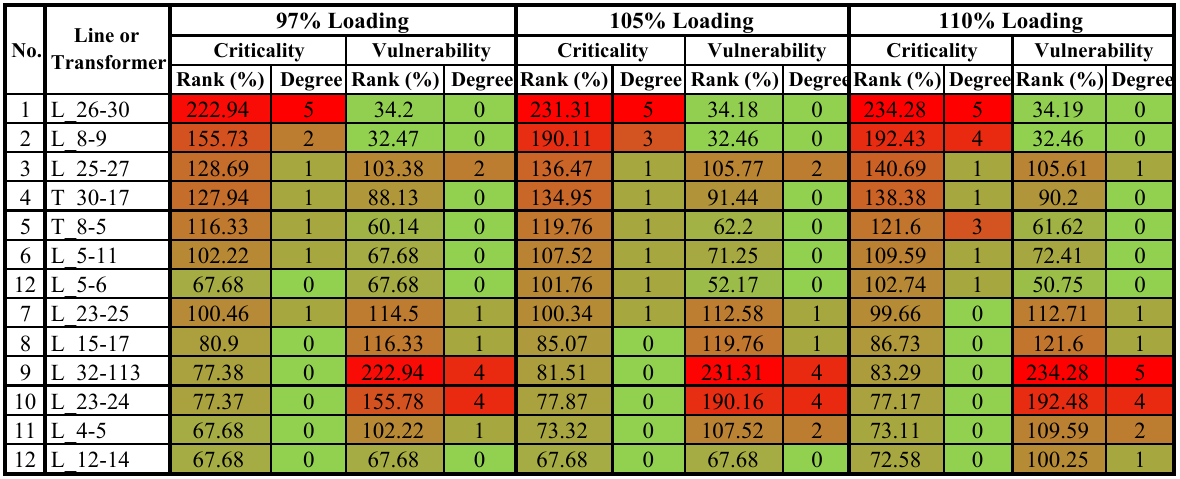}
\end{table*}

\section{Case Studies}

The case studies are conducted on the IEEE 118-bus test system. We use different loading scenarios at 97\%, 105\%, 106\%, and 110\% of the system's peak load to generate various stress levels. The nodal loads are uniformly adjusted for all the cases, except for the 106\% loaded case, where the demand is increased only on select buses (16\% in West End (40) and 105\% in S. Tiffin (41)). Then, we run an AC optimal power flow for each loading scenario to obtain AC feasible base case solutions for the analysis. These solutions are fed to the PowerWorld Simulator to calculate stress metrics and examine transmission switching impacts.

We assume that all lines have their contingency limits at 120\% of the normal limits, which can be used for a limited duration of 4 hours. We further assume the emergency limits to be at 135\% of the normal line limits, which can be used up to 15 minutes~\cite{nerc2}. We acknowledge that the contingency and emergency limits are not necessarily always scaled uniformly to the normal limits; however, we make this assumption to simplify the paper's analysis. We further acknowledge that these limits may change depending on the weather or loading scenarios; again, we have neglected such details to simplify the analysis.

Table~\ref{fig:table_97} presents the stress analysis with different loading scenarios, as described earlier. The purpose of these stress tables is to demonstrate how loading with different patterns can affect the system stress metrics. Generally, the stress increases with the loading; however, the 106\% loaded case, with the non-uniform increase in the nodal loads, is atypically stressed, even beyond the 110\% loaded case. This demonstrates that the distribution of the load has a significant impact on system stress. We pick this atypically stressed case to demonstrate the benefits of preventive transmission switching.

\begin{figure*}[h!]
	\centering
	\includegraphics[width=4.7in]{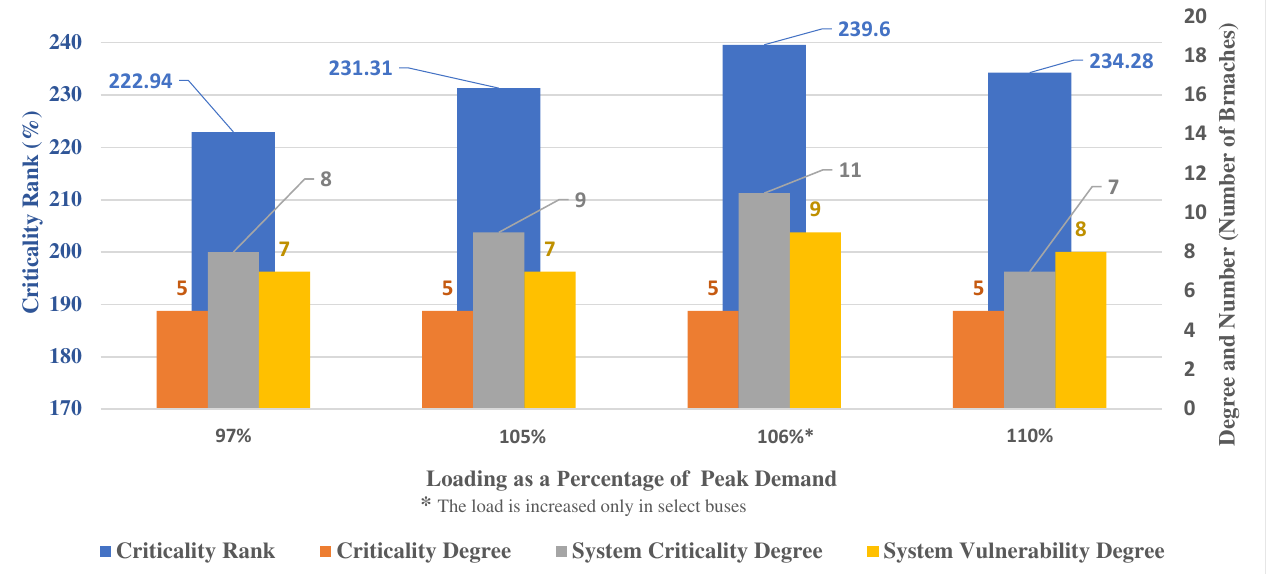}
	\caption{Comparison of the stress on the IEEE 118-bus test system in terms of criticality rank, degree, and system degree under different loading scenarios.} \label{fig:loadstresscomparision}
\end{figure*}

\begin{table}
	\centering
	\caption{The criticality stress metrics for the 106\% loaded IEEE 118 test case before and after the implementation of a single preventive switching action (17 Sorenson - 113 Deer Crk). } \label{fig:table_106_PTS}
	\includegraphics[width=2.7in]{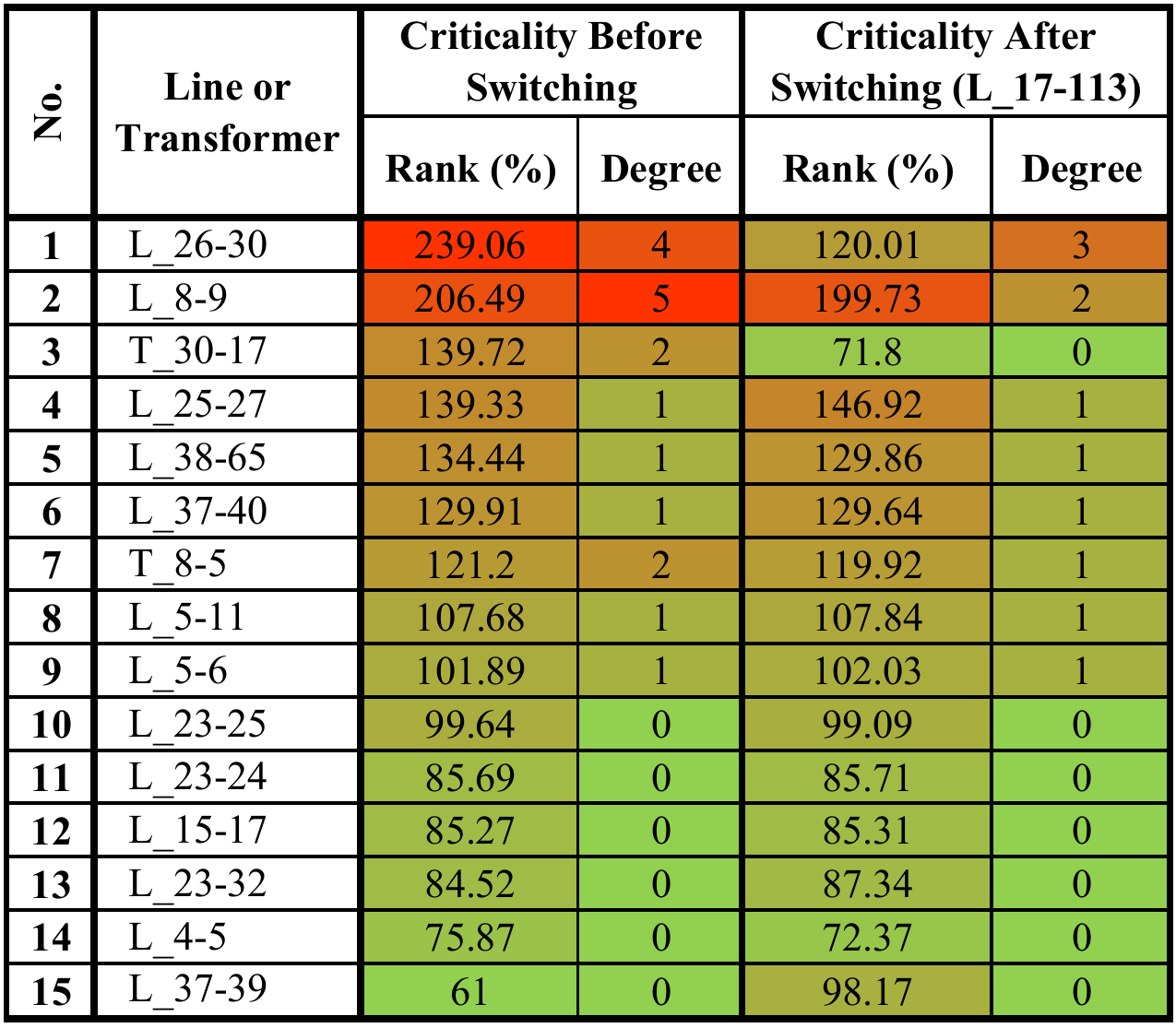}
\end{table}

\begin{table*}[t]
	\centering
	\caption{Various stress measurements for the system under different loading scenarios, including the 106\% loaded case after implementation of one preventive transmission switching action.}
	\begin{tabular}{|c|c|c|c|c|c|c|c|c|}
		\hline
\hline
  Loading \% of & \multirow{2}{*}{$V^{rank}$} &\multirow{2}{*}{$V^{degree}$} &\multirow{2}{*}{$V_{N}$} & \multirow{2}{*}{$C^{rank}$} &\multirow{2}{*}{$C^{degree}$} &\multirow{2}{*}{$C_{N}$}  &Emergency limit &Contingency limit \\ 
  peak demand& &  &  &  &  &  & Violation & Violation  \\ \hline

97\%	&222.94\%	&5 	&8 	&222.94\%	&4  &6	&2 &4 \\ \hline
105\%	&231.31\%	&5 	&8 	&231.31\%	&4 	&6	&3 &4 \\ \hline
$106$\%	&239.06\%	&5 	&9 	&239.06\%	&5 	&9	&4 &7 \\ \hline
$106$\% (PTS)	&199.73\%	&3 	&8 	&199.73\%	&4 	&7	&2 &4 \\ \hline
110\% &234.28\%	&5 	&7 	&234.28\%	&5 	&7	&3 &4 \\ \hline
 
 \hline
   \end{tabular}
\label{table:result}
\end{table*}
\begin{figure*}[h!]
	\centering
	\includegraphics[width=4.8in]{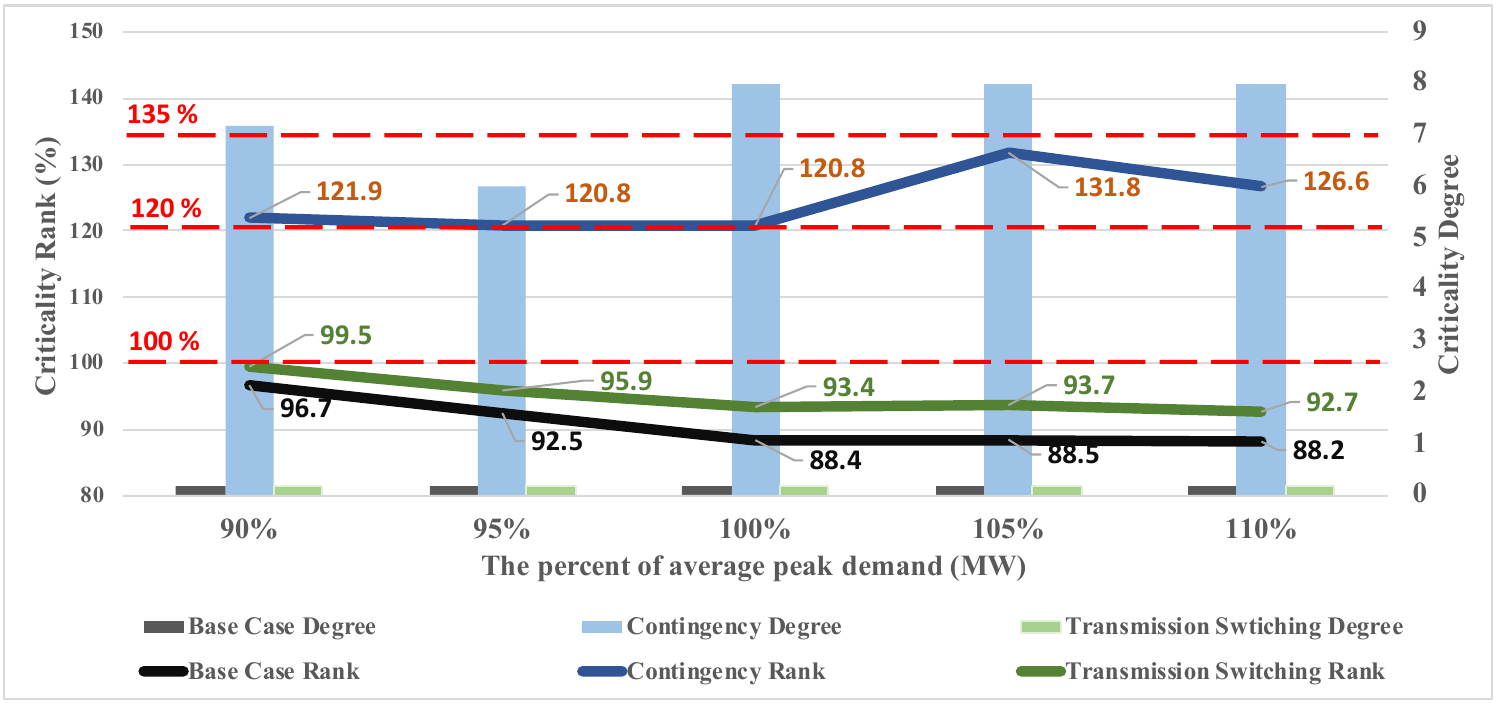}
	\caption{Comparison of system stress in terms of criticality rank and degree under different loading scenarios for the base case, contingency case (an outage of 8 Olive - 5 Olive Transformer), and post corrective transmission switching of 15 Ft. Wayne - 17 Sorenson.} \label{fig:CTS}
\end{figure*}

\subsection{Preventive Transmission Switching}
\begin{figure}[h!]
	\centering
	\includegraphics[width=3.4in]{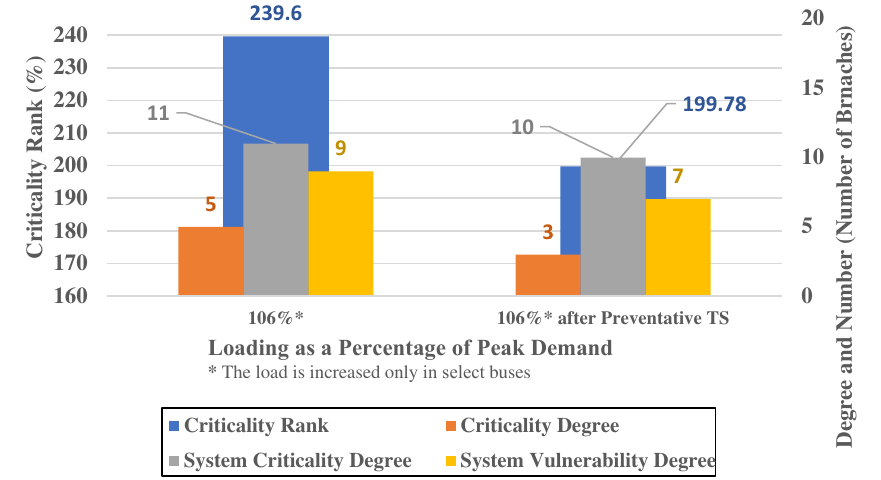}
	\caption{The stress comparison of pre and post transmission switching for the 106\% loaded IEEE 118 bus test case. The rank and degree in this chart represent the single highest value for the system.} \label{fig:prevetativeTS}
\end{figure}

\begin{figure}[h!]
	\centering
	\includegraphics[width=3.2in]{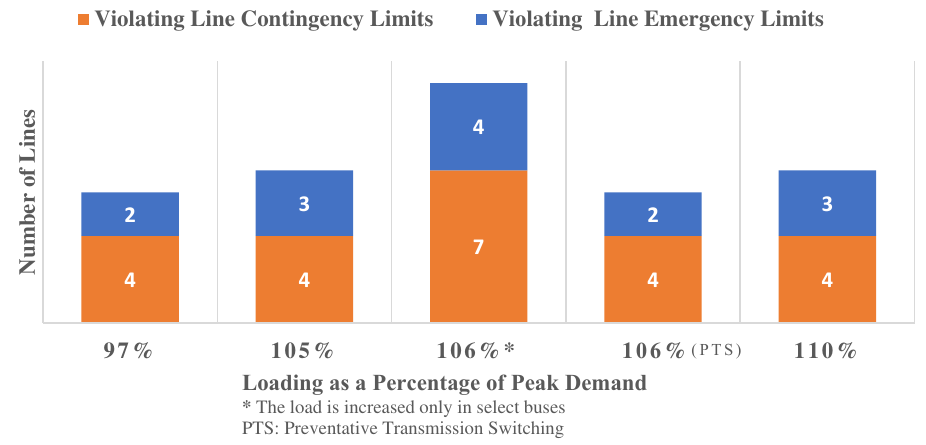}
	\caption{The plot compares the number of lines, violating both emergency and contingency limits under different loading scenarios, including the post preventive transmission switching.} \label{fig:lineviolation}
\end{figure}

Figure~\ref{fig:loadstresscomparision} shows the various stress metrics, comparing the stress on the IEEE 118-bus case system in terms of maximum criticality rank, maximum degree, and system degree under different loading scenarios. As can be seen, the stress for the case with 106\% loading is atypically high. Table~\ref{fig:table_106_PTS} shows the stress on the same case before and after a preventive switching action is implemented, where 17 Sorenson - 113 Deer Crk line is opened, but the generation dispatch is not changed. Full stress analysis is performed for pre- and post- transmission switching, and the stress comparison for this case is shown in Fig.~\ref{fig:prevetativeTS}. The Switching Stress Relief indices calculated for this loading condition as $SSR_d^{17-113}=6$ and $SSR_r^{17-113}=199.86$. The plot, comparing the number of lines loaded above both emergency and contingency limits, under different loading patterns, including after preventive transmission switching, is shown in Fig.~\ref{fig:lineviolation}. The results of different stress metrics parameters are tabulated in Table~\ref{table:result}. The results clearly show that a single transmission switching action can substantially reduce the system stress and avoid a potential cascading failure event.

\subsection{Corrective Transmission Switching}
One of the highly critical non-radial contingencies, identified by PowerWorld Simulator, is 8 Olive - 5 Olive transformer. For this contingency, the system's stress is such that at least one of the in-service lines exceeds the contingency limit. Therefore, the operator has about 15 minutes to address this issue, or another line will be tripped. The algorithm developed in this paper suggests switching of 15 FtWayne - 17 Sorenson line as a corrective action to reduce the system stress to an acceptable level.

Fig.~\ref{fig:CTS} compares the stress on the system in terms of criticality rank and degree under different loading scenarios for the base case, contingency (8 Olive - 5 Olive transformer), and corrective transmission switching of 15 FtWayne - 17 Sorenson line for the IEEE 118-bus test case. The results confirm the effectiveness of corrective transmission switching in reducing the post-contingency system stress, close to the normal operation levels.

\section{Conclusion}
System stress metrics are recently developed to provide insights into the susceptibility of the system to cascading failures. The metrics include measures of the criticality of contingencies and vulnerability of the transmission elements to overloads after contingencies. Building upon those metrics, this paper introduced two new metrics to measure the system's criticality and vulnerability. All of these metrics can be quickly calculated via the outputs of the contingency analysis tool or through LODFs. Furthermore, the paper investigated the possibility of employing transmission switching, both as a preventive and corrective measure, to reduce the system stress. LODF sensitivities were used to generate a relatively small subset of quality switching candidates and achieve computational tractability for the transmission switching algorithm. Those candidates were then tested for effectiveness until a practical solution was found or the list was depleted. The simulation studies on the IEEE 118-bus system confirmed the efficacy of the method. A single transmission switching action was able to reduce the system stress to the normal levels substantially. This implies that system operators should look at transmission switching as a potentially useful tool to prevent cascading failures when the system is atypically stressed. Other alternative actions, such as generation redispatch, are substantially more expensive and may also not be available when the system is highly stressed.



%



\section*{Acknowledgment}

The authors would like to thank Dr. Hyde Merrill for his careful review and detailed feedback. We also thank Dr.  Marc  Bodson for providing us with the PowerWorld Simulator license.

\ifCLASSOPTIONcaptionsoff
  \newpage
\fi



%
\bibliographystyle{IEEEtran}
\bibliography{IEEEabrv,ref}

\end{document}